\documentclass[11pt]{article}
\usepackage{fancyhdr}
\usepackage{isomath}
\usepackage{amsmath}
\usepackage{amsbsy}
\usepackage{amssymb}
\usepackage{amscd}
\usepackage{amsfonts}
\usepackage{graphicx,color}
\usepackage{verbatim}
\usepackage{euscript}
\usepackage{alltt}
\usepackage{stmaryrd}
\usepackage{subfigure}

\usepackage{amsmath}
\usepackage{amsbsy}
\usepackage{amssymb}
\usepackage{amscd}
\usepackage{amsfonts}

\newcommand{\bfa}{{\mathbold a}}
\newcommand{\bfb}{{\mathbold b}}

\newcommand{\bfd}{{\mathbold d}}

\newcommand{\bff}{{\mathbold f}}

\newcommand{\bfn}{{\mathbold n}}

\newcommand{\bfp}{{\mathbold p}}

\newcommand{\bft}{{\mathbold t}}
\newcommand{\bfu}{{\mathbold u}}

\newcommand{\bfC}{{\mathbold C}}

\newcommand{\bfS}{{\mathbold S}}
\newcommand{\bfT}{{\mathbold T}}
\newcommand{\bfU}{{\mathbold U}}
\newcommand{\bfV}{{\mathbold V}}

\newcommand{\bfX}{{\mathbold X}}

\newcommand{\beq}{\begin{equation}}
\newcommand{\eeq}{\end{equation}}
\newcommand{\beqs}{\begin{eqnarray}}
\newcommand{\eeqs}{\end{eqnarray}}
\newcommand{\beql}{\begin{equation} \label}

\newcommand{\del}{\partial}

\newcommand{\bfalpha}{\mathbold{\alpha}}

\newcommand{\bfPi}{\mathbold{\Pi}}

\usepackage[margin=1in]{geometry}
\date{March 4, 2016}

\begin{document}
\title{A microscopic continuum model for defect dynamics in metallic glasses}
\author{Amit Acharya\footnote{
Dept. of Civil \& Environmental Engineering, CMU, email: acharyaamit@cmu.edu}$\ $ 
and Michael Widom\footnote{
Dept. of Physics, CMU, email: widom@andrew.cmu.edu}\\\\
Carnegie Mellon University, Pittsburgh, PA 15213}

\maketitle

\begin{abstract}
\noindent Motivated by results of the topological theory of glasses accounting for geometric frustration, we develop the simplest possible continuum mechanical model of defect dynamics in metallic glasses that accounts for topological, energetic, and kinetic ideas. The model is aimed towards the development of a microscopic understanding of the plasticity of such materials. We discuss the expected predictive capabilities of the model vis-a-vis some observed physical behaviors of metallic glasses.

\end{abstract}

\section{Introduction}
Despite their technological relevance, there does not exist a fundamental characterization of the process of plastic deformation in a metallic glass, relating its structure and `defects' to deformation \cite{schuh2007mechanical, greer2013shear}; indeed, whether any structural defect in a glass can be characterized at all seems to be an open question. In many ways this is surprising as there exists a beautiful body of work in the physics literature based on topological ideas related to geometric frustration and the so-called polytope model of glass that does rationalize defects in glass structure that are linked to atomic configurations (of at least mono-atomic materials interacting by pair potentials). Glasses have also been vigorously studied from the mechanics of materials and materials science perspectives, but without making connection to the physics literature related to geometric frustration. Currently, there is no theory available to study the far-from-equilibrium mechanical response of bulk metallic glasses that gives equal importance to topological, energetic, and kinetic
ideas. As mentioned, a primary hindrance has been the unequivocal identification of a structural defect as the essential carrier of inelastic deformation in metallic glasses \cite{srolovitz1981structural}. The concept of a shear transformation zone originated by Argon \cite{argon1979plastic} and developed by Falk and Langer \cite{falk1998dynamics} comes close, but it is not identifiable as a rigorous structural state variable from atomic configurations of glass-forming alloys. The work of Spaepen \cite{spaepen1977microscopic} related to ‘free-volume’ is the other advance in the direction of characterizing a structural state variable for glasses. Quite surprisingly, the ideas of Nelson \cite{nelson1983order}, making a direct connection between `glass dislocations' consisting of disclination-dipoles and atomic configurations, have not been considered in the vast literature on attempts to understand structure-deformation connections in metallic glasses \cite{falk1998dynamics, takeuchi2011atomistic, schuh2007mechanical, greer2013shear, anand2005theory}. Our goal in this paper is to address this gap with the hope that it may lend complementary insight into the modeling of the mechanical behavior of metallic glasses and amorphous materials.

Specifically, we will model the metallic glass as a sea of geometrically frustrated regular tetrahedra arranged predominantly in clusters of five around static backbones of 5-fold disclination lines. This sea is thought of as punctuated by mobile 4 and 6-fold disclination dipole lines that can be interpreted as dislocation lines (with spread out cores) in the medium, following the topological theory of defects \cite{nelson1983order}. There appears to be evidence that regions of non-pentagonal packing seem to suffer the most plasticity for amorphous materials \cite{takeuchi2011atomistic, ding2014soft}. We combine these insights based essentially on homotopy theory describing possible lowest-energy static states of the glass with a partial differential equation based model of dislocation dynamics. The result is a model for dissipative defect dynamics in metallic glasses and similar amorphous materials. With no further assumptions beyond this rigorous kinematics and the simplest linear kinetic assumptions arising from enforcing the second law of thermodynamics, we show that the proposed model provides plausible explanations for
\begin{itemize}
\item the origin of a stochastic, observed internal stress field
\item dilatancy in observed plastic flow of metallic glasses
\item pressure dependence of observed plastic flow
\item threshold behavior in the motion of these glass dislocations in response to stress
\item structure and dynamics of deformation localization in the form of shear bands; in particular their
longitudinal propagation taking full account of the effects of material inertia.
\end{itemize}

The outline of the paper is as follows: in Section \ref{sec:lit_rev} we briefly review a few elements of the vast literature on the physics and engineering of amorphous solids, including a summary of Nelson's proposal \cite{nelson1983order} for modeling the kinematics of glasses. Section \ref{sec:main_idea} describes  the adaptation of Nelson's idea to a flat, 3-d space treatment, showing how, in the first approximation, `glass-dislocations' may be rationalized. We substantiate these ideas with quantum mechanical electronic structure  calculations showing these defects.  With this kinematic basis, in Section \ref{sec:disloc_theory} we adapt a continuum model for dislocation dynamics \cite{acharya2001model, acharya2010new, zhang2015single}, but now involving disclinations as well for goals particular to this application,  and show some of its implications related to the `predictions' mentioned above.

\section{Brief review of the literature}\label{sec:lit_rev}
\begin{figure}
\centering
\includegraphics[width=4.5in, height=4.0in]{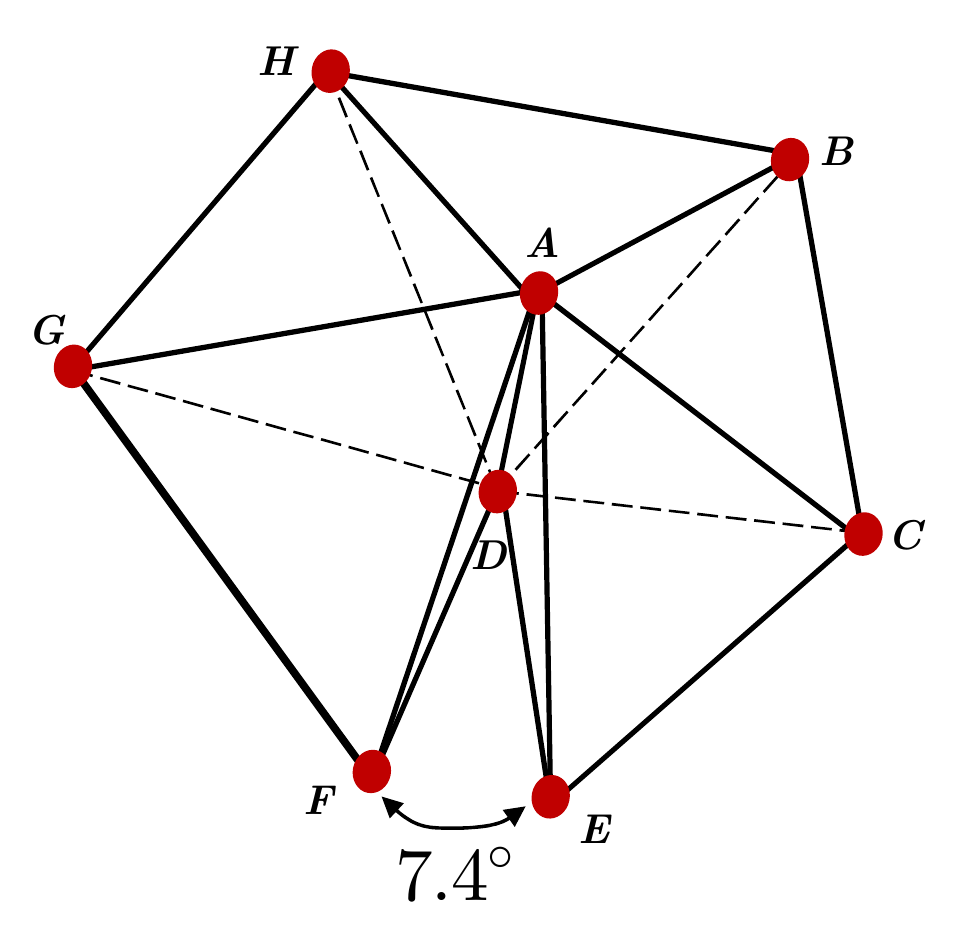}
\caption{Schematic of geometric frustration arising from tiling space with regular tetrahedra.}
\label{fig:frustration}
\end{figure}

\begin{figure}
\centering
\includegraphics[width=6.0in, height=3.0in]{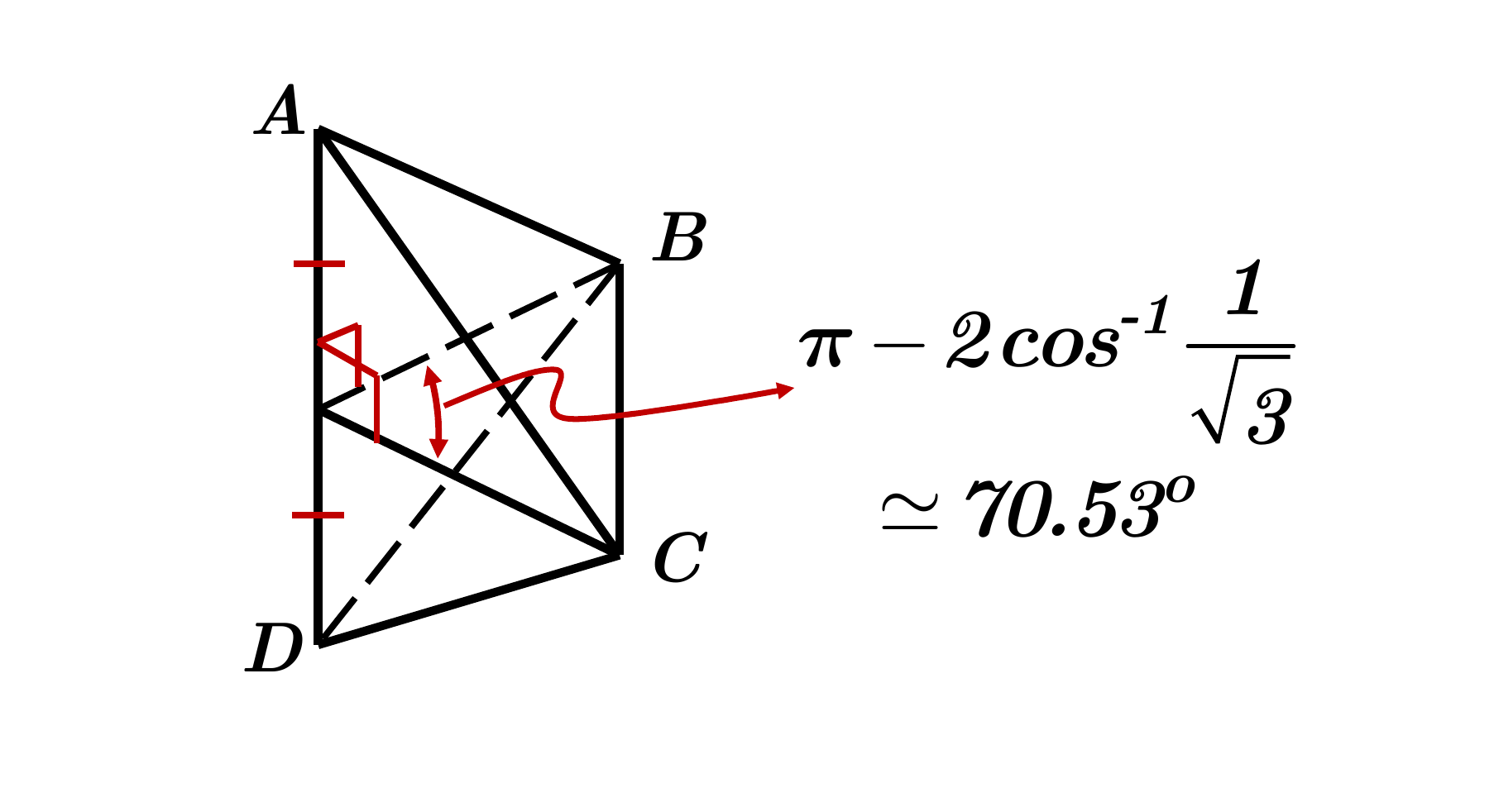}
\caption{Geometry of a regular tetrahedron.}
\label{fig:tet_geom}
\end{figure}

The fundamental idea in understanding the mechanics of glasses is geometric frustration \cite{kleman2008disclinations}. This refers to the fact that \emph{regular} tetrahedra, all with equal edge-lengths, cannot tile space. The easiest way to see this is to visualize a set of five regular tetrahedra sharing a common edge as in Figure \ref{fig:frustration}.
The dihedral gap-angle is calculated based on the geometry of a regular tetrahedron shown in Figure \ref{fig:tet_geom}.
One considers tetrahedra as the basic structural building block since it is easy to see \cite{nelson1983order} that for atoms interacting by pair potentials, a regular tetrahedron of four atoms has to be a ground state. Putting five such tetrahedra without a gap as in Fig. \ref{fig:frustration} necessarily means an atomic assembly under stress. In computer simulations of glasses, such pentagonal order is found to be predominant \cite{kleman1979tentative, takeuchi2011atomistic}. Putting together two such stressed assemblies of 7 atoms each sharing one central atom (atom D in Fig. \ref{fig:frustration}), where one assembly is rotated with respect to the other about the central spine (the extension of the line AD in Fig. \ref{fig:frustration}) creates a 13-atom icosahedron with 20 faces.

In order to explain the observations of Turnbull \cite{turnbull1981metastable} on the supercooling of liquids, Frank \cite{frank1952supercooling}  rationalized that icosahedral order should be expected in small clusters as it is energetically favorable in Lennard-Jones materials; for the same number of atoms, an icosahedron has higher binding energy than the fcc and the hcp structures. Later, Honeycutt and Anderson \cite{honeycutt1987molecular} observed, using zero-temperature molecular dynamics simulations, that Lennard-Jones clusters below $5000$ atoms preferred the Mackay icosahedral state over the fcc structure. The attainment of icosahedral order in metallic glasses and supercooled liquids has since been substantiated by experiment in \cite{liu2013systematic, schenk2002icosahedral, di2003there} and in simulation in \cite{sheng2006atomic, ganesh2006signature}. Interestingly, when some of the structures in \cite{ganesh2006signature} were relaxed to obtain their ground states, many crystallized, a fact not inconsistent with the idea that extended assemblies of atoms of Cu, driven by the periodic boundary conditions of the simulations, can prefer a crystalline ground state.

Noting that random packings of hard spheres show pentagonal order, e.g. a prevalence of rings of 5-bonds, arising from the energetic preference for tetrahedra as mentioned above, Kleman and Sadoc \cite{kleman1979tentative} argue that defects have to be present in extended assemblies of such atoms and proceed to classify the possible types of defects. They ask if there are natural, geometric ways of seeking such tetrahedral packings with associated defects in 3-d space. They go on to study the $\{3,3,5\}$ regular polytope \cite{coxeter1973regular} that tiles perfectly the 3-d sphere (that could be viewed as being embedded in 4-d Euclidean space), each vertex of which is the center of an icosahedron. Squashed into 3-d space, such a polytope generates local tetrahedral ordering and \emph{disclination} line defects \cite{sadoc1982order}. Rivier and Duffy \cite{rivier1982line} show that rings with odd-numbered bonds in a `continuous random network' have to be closed loops or end at boundaries and that they cost energy; it is suggested that these lines be viewed as rotational disclination lines.

Nelson \cite{nelson1983order} suggests that supercooled liquids and metallic glasses be viewed as ``icosahedral liquid crystals with, however, a global excess of defects with a particular sign'' arising from the curvature incommensurability of the 3-d sphere and our ambient 3-d space. Very importantly, Nelson gives a prescription based on the Voronoi construction for tesselating an atomic assembly and identifying defect lines. While Nelson thinks about disclinations, his idea is different from Rivier's. Noting the preponderance of 5-ring disclination lines in a glass, he assumes material filled with such lines as the background in which dipoles of 4-ring and 6-ring disclination lines act as the predominant composite \emph{dislocation} defects. However, because of the curvature incommensurability, there necessarily has to be an excess of 6-rings and he develops a quantitative model for the average number of tetrahedra around any bond to be expected in amorphous assemblies. This number comes out to be slightly in excess of 5, matching an estimate of Coxeter \cite{coxeter1973regular} for a `statistical honeycomb' model. All of the above theoretical works are based in the topological theory of defects. Sachdev and Nelson \cite{sachdev1984theory} utilized a Landau theory based on frustrated icosahedral order to predict the structure factor (experimental signature) of metallic glass, as is also discussed in Nelson and Widom \cite{nelson1984symmetry}.  A parallel development is the work of Sethna \cite{sethna1983frustration, sethna1985frustration} drawing a parallel between metallic glasses and cholosteric blue phases.

The mechanics of materials literature on the modeling of glasses begins with the \emph{free-volume} model due to Spaepen \cite{spaepen1977microscopic} followed by Argon \cite{argon1979plastic} who defines a \emph{shear transforming zone} for the first time as deformation where ``the transformation is in a narrow disk shaped region and resembles closely the nucleation of a dislocation loop." A phenomenological model with some success has been the one of Anand and co-workers \cite{anand2005theory} which is a Coulomb-Mohr `crystal' plasticity model including dilatational plasticity based on free-volume evolution, whose slip systems are appropriately designed. A different line of phenomenological reasoning based on `disorder' as state variable and corresponding  `effective temperature' thermodynamics is presented in \cite{bouchbinder2009nonequilibrium, kamrin2014two}.  Comprehensive reviews are to be found in \cite{schuh2007mechanical} and \cite{greer2013shear}, the latter focused on shear-banding in metallic glasses.

Our goal in this paper is to develop a mesoscopic theory of deformation of metallic glasses that connects to their atomic structure and the line defects in the structure as mentioned above.

\section{Fundamental motivation for elasticity and defect kinematics in proposed model}\label{sec:main_idea}
\begin{figure}
\centering
\includegraphics[width=5.5in, height=2.0in]{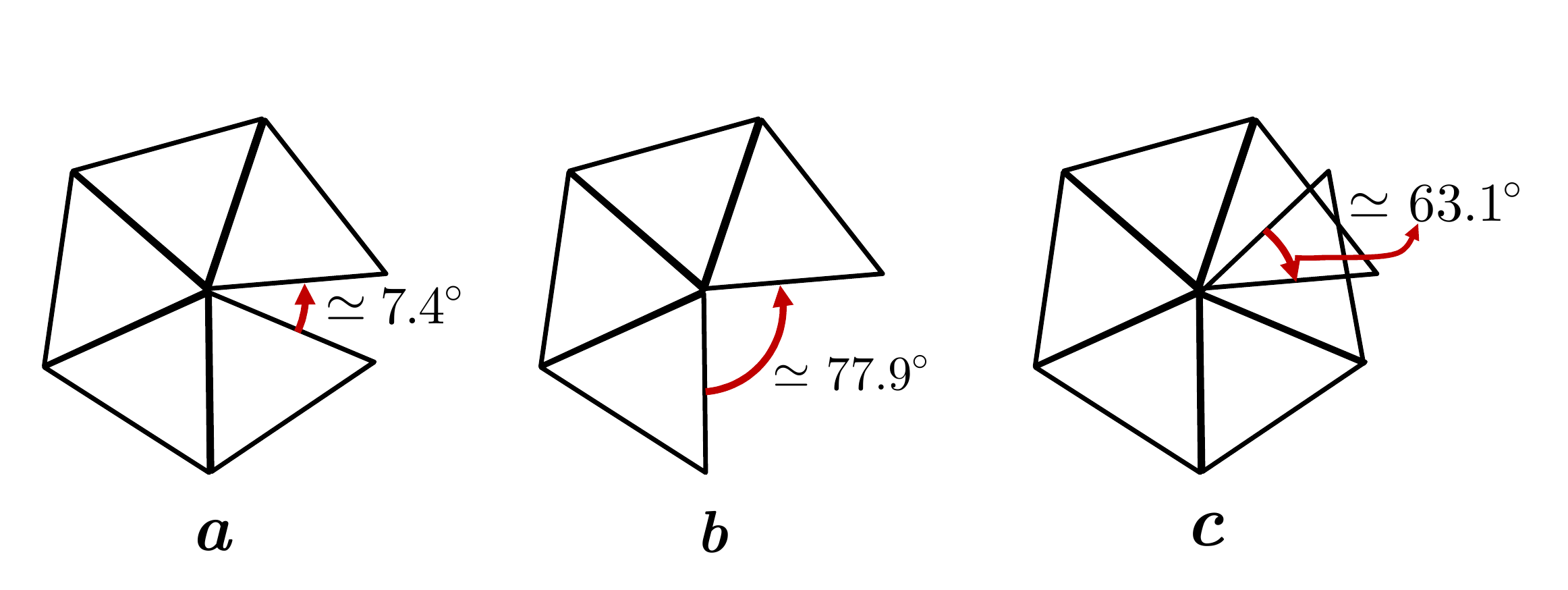}
\caption{Stress-free states of 5, 4, and 6 bi-pyramids giving rise to 5, 4, 6-fold disclination lines.}
\label{fig:bi-pyramids}
\end{figure}

The assembly of five tetrahedra shown in Fig. \ref{fig:frustration} will be referred to as the stress-free state of a \emph{5 bi-pyramid} (motivated by the side-view along rays in the plane perpendicular to the bond AD). Strictly speaking, this state is physically unrealizable as there is only one atom in place of the two at F and E, and the whole structure is therefore strained. As shown, it is to be understood that the atoms E and F do not interact. Figure \ref{fig:bi-pyramids} shows the plan view of the stress-free states of 5, 4 and 6 bi-pyramids. Their corresponding stressed states arise by closing or eliminating the shown angular gaps and letting the body relax. The \emph{axes of these bipyramids in the stressed state are defined as 5, 4 and 6-fold disclination lines} in the glass structure, with the understanding that the structure is continued into the page with the axis as a curve running from one boundary of the body to another or as a closed loop within the body; along the axes the 5,4,6 rings twist from one to the next to maintain overall tetrahedral coordination (much as in the formation of an icosahedron consisting of two 5-rings).
Clearly, the 5 and the 4-fold disclination lines cause tensile stress while the 6 fold disclination causes compressive stress in the body. Equally clearly, the 6-fold disclination causes less energy to be stored in the material than the 4-fold one (as the angular adjustment required is less). Thus, in rough terms based purely on considerations of minimization of stored energy in the body, it may be expected that 6-fold disclination lines would be in excess of the 4-fold ones, contained within a background of a dense collection of 5-fold disclinations that cannot be avoided on grounds of geometric frustration. However, it seems natural that if a 4-fold disclination and a 6-fold disclination were to be brought in close proximity then their compressive and tensile stress fields would counterbalance each other to produce much less stored energy. We shall call such a combination of two disclination lines of opposite character as a \emph{glass dislocation} or simply a dislocation in our context. In \cite{nelson1983order, nelson1989polytetrahedral} reasoning is provided for the existence of such combined defect lines on the basis of the homotopy theory of defects as well as atomic configurations in an icosahedral medium.

The complete homotopy theory of topological defects in icosahedral media is presented in \cite{nelson1984symmetry}. There it is
shown that a generic line defect can be generated as a combination of left and right-handed screw operations. Pure
disclinations are combinations with no net translation, while pure dislocations have no net rotation, for example a pair consisting of a 6-fold and a 4-fold disclination (referred to in the following as a $6/4$ pair). When line defects cross they generate an umbilical defect line joining them. Thus to model a pure dislocation it is desirable to let the $6/4$ pair run parallel to each other. 

\begin{figure}
\centering
\includegraphics[width=4.0in, height=4.0in]{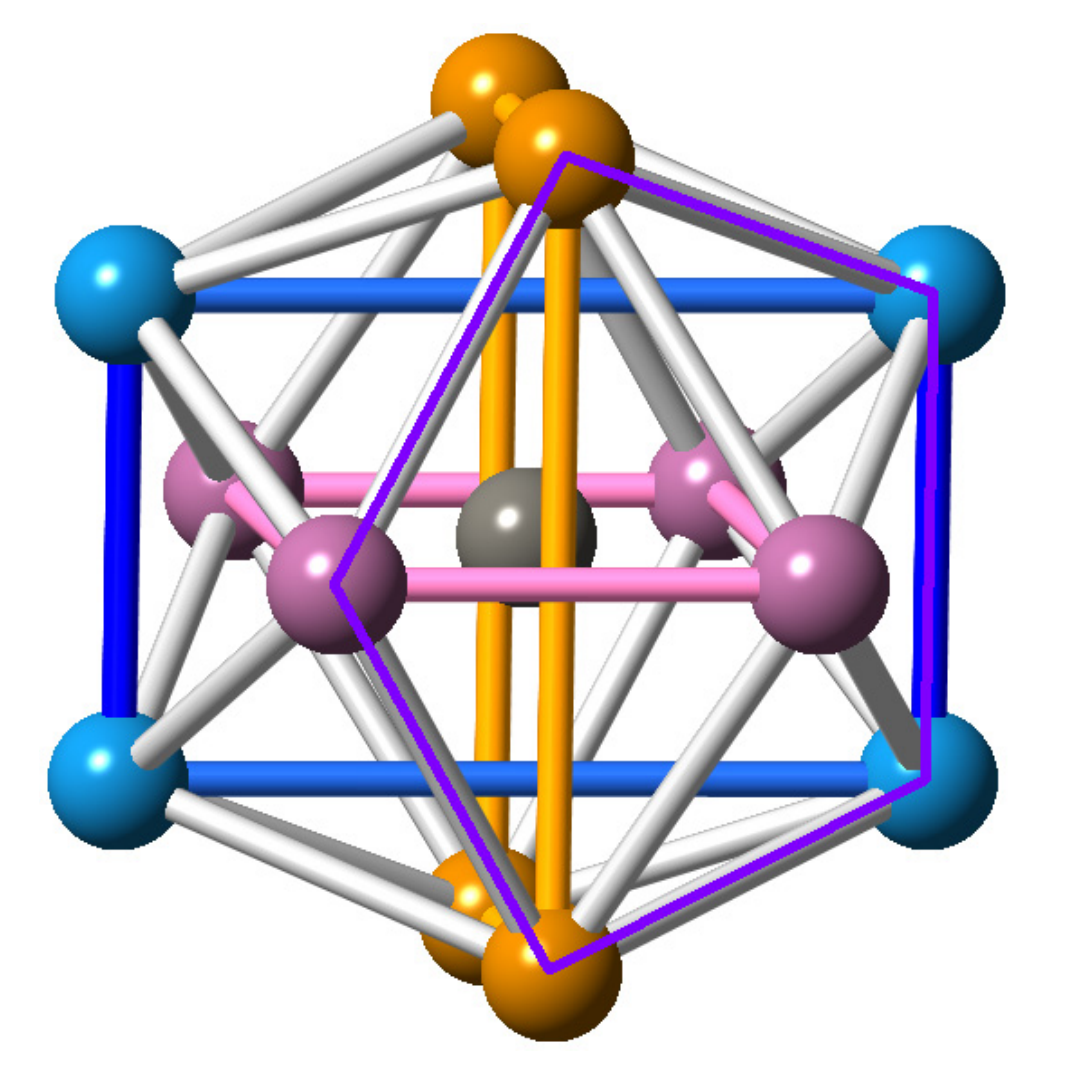}
\caption{$Z12$, a perfect icosahedral 13-atom cluster; the central atom has $Z=12$ coordination.}
\label{fig:Z12}
\end{figure}

We now exhibit specific atomic realizations of these disclinations and resulting dislocations for the sake of illustration.
Consider first the ideal 13-atom icosahedral cluster (Fig. \ref{fig:Z12}).
We denote this structure as $Z12$, because the central
atom has $Z = 12$ neighbors \cite{nelson1983order, frank1958complex}.
Every bond from the central atom to a surface atom is five-coordinated, i.e. surrounded by a ring of 5 atoms that are
simultaneous neighbors of the central atom and the surface atom. This figure also shows how the 12 surface atoms can
be grouped into three sets of four, with each set occupying the vertices of a golden rectangle and these rectangles being
mutually perpendicular.

\begin{figure}
\centering
\includegraphics[width=4.0in, height=3.7in]{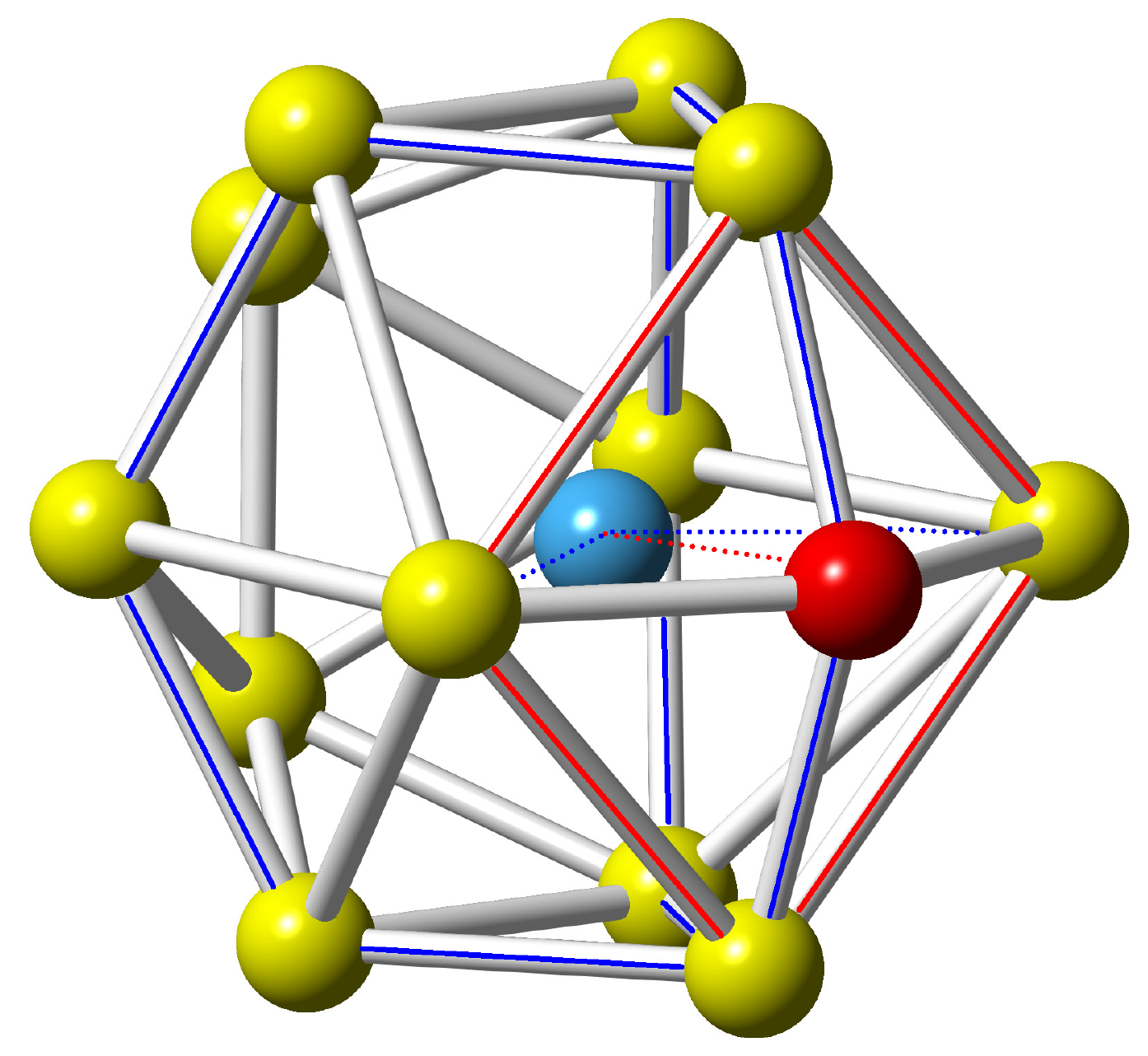}
\caption{$Z13$ cluster with $14$ atoms.}
\label{fig:Z13}
\end{figure}

Now we introduce disclinations. As shown by Nelson \cite{nelson1983order}, the $Z13$ cluster (Fig. \ref{fig:Z13})
formed by inserting an interstitial atom into the surface of the $Z12$ possesses one 4-fold and two 6-fold disclinations.
The 4-fold disclination lies along a bond from the center to surface that is four-coordinated, i.e. surrounded by a
ring of four simultaneous neighbors. The 6-fold disclinations are siz-coordinated and surrounded by a ring of six
simultaneous neighbors. The atomic coordinates in this figure were obtained by relaxation of an atomic cluster
containing a Tantalum atom at the center, twelve Aluminum atoms and one Iron atom on the surface.
Forces were calculated from quantum mechanical first principles using electronic density functional theory
\cite{kresse1996}.

\begin{figure}
\centering
\includegraphics[width=6.0in, height=6.0in]{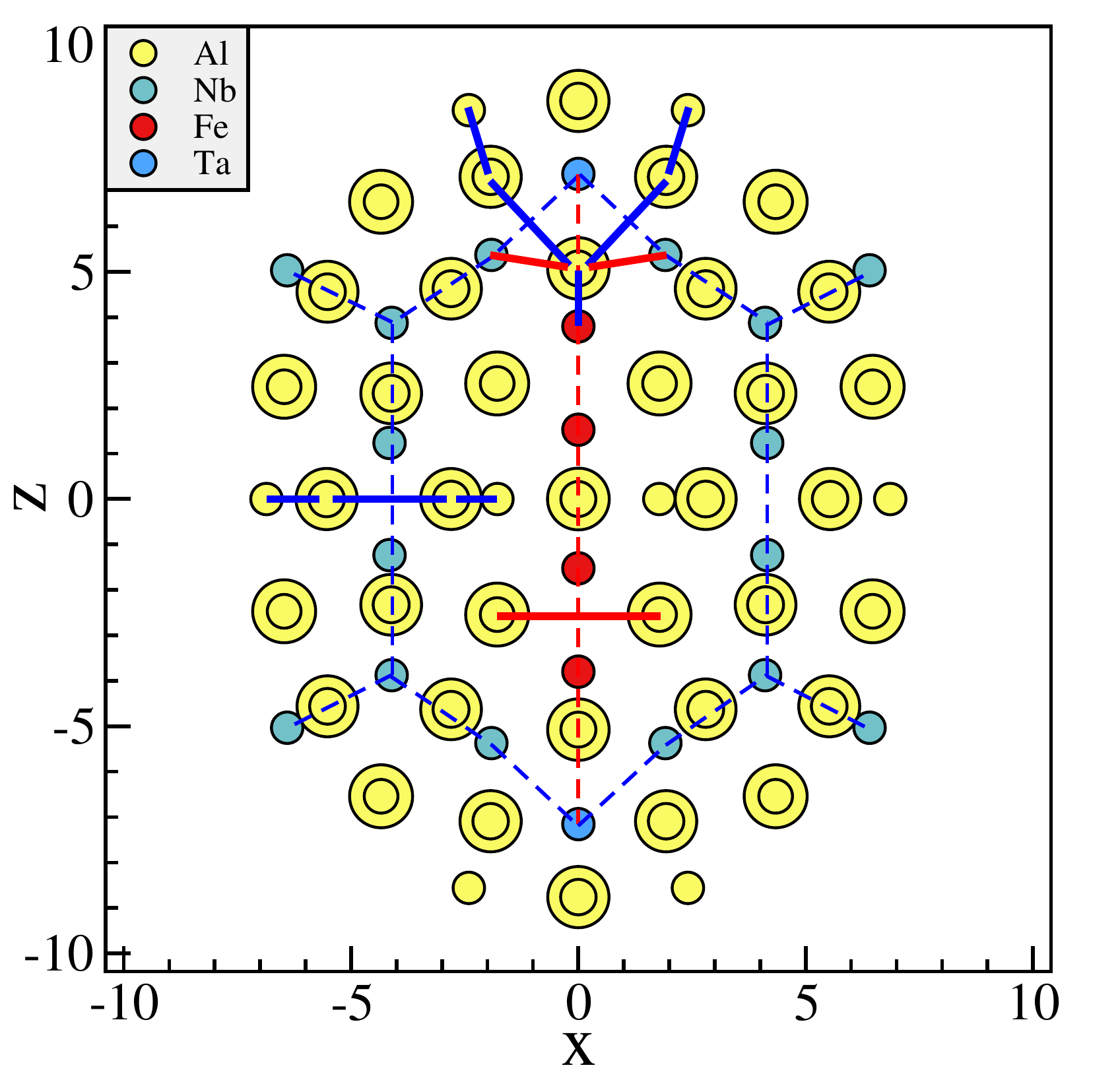}
\caption{Example disclination network relaxed using first principles electronic density functional theory. Chemical species are color coded. Sizes indicate position perpendicular to plane with large below and small above. Dashed line segments are disclination lines, solid  segments are bonds. Blue corresponds to 6-fold and red is 4-fold (for both disclination lines and rings surrounding them). Ta atoms are $Z13$ centers terminating two 6-fold and one 4-fold disclinations. Nb atoms are $Z14$ centers with 6-fold disclinations passing through, except for triple junctions, which are $Z15$ centers. All Fe atoms are $Z10$ centers with 4-fold disclinations passing through. For the $Z13$ at the top, some bonds and disclination line segments visible in this view are color matched with corresponding entities in Fig. \ref{fig:Z13}.}
\label{fig:x-z-view}
\end{figure}

\begin{figure}
\centering
\subfigure[]{
\includegraphics[width=4.0in, height=4.0in]{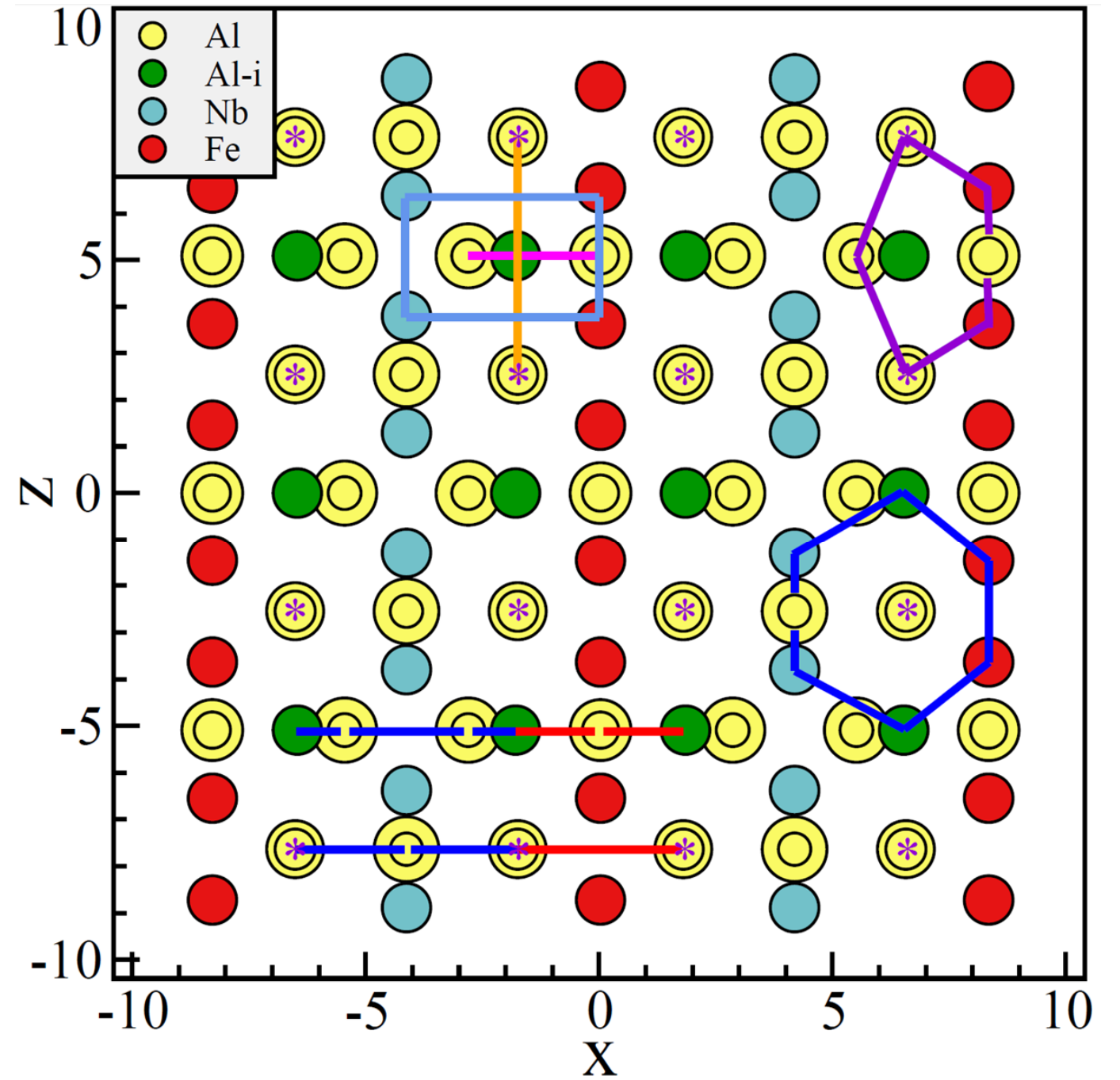}
\label{fig:2a_final}}\\
\subfigure[]{
\includegraphics[width=4.5in, height = 2.5 in]{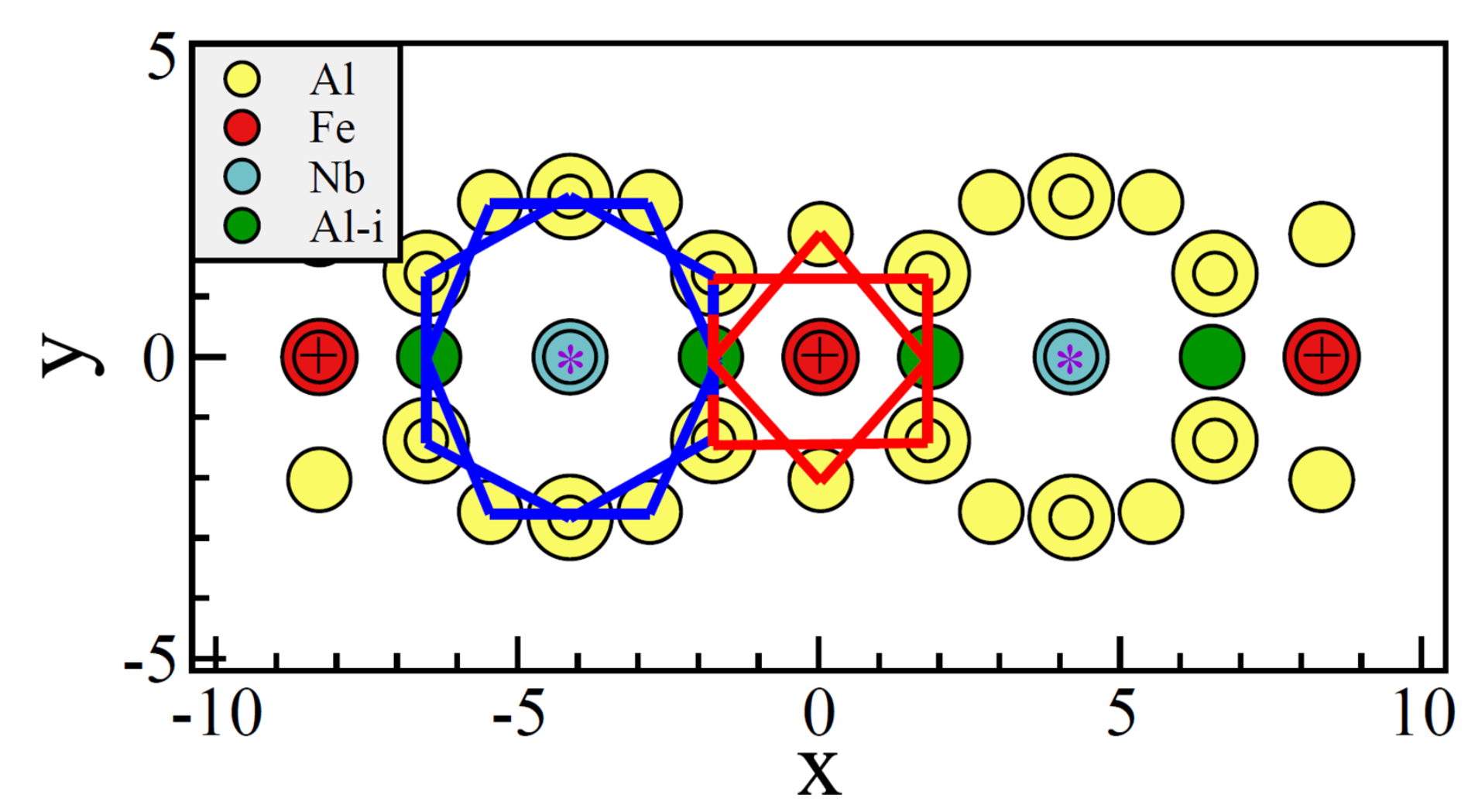}
\label{fig:2b_final}}
\caption{Periodic extension of parallel $6/4/6$ disclination lines in the $x-z$ plane.
(a) View of $x-z$ plane. Icosahedrally coordinated
$Z12$ Al atom (marked Al-i in green). Neighbors are connected by three orthogonal rectangles (two viewed edge-on) as
indicated in light blue, magenta and orange at top (color matched with Fig. \ref{fig:Z12}). 6-fold and 4-fold rings are indicated in blue and red, respectively, at
bottom. Additional 6-fold disclination lines run perpendicular to the plane at locations marked with asterisks. One 6-fold
ring is shown in blue at right. A 5-fold ring of a $Z12$ is shown at top right in purple (color matched with Fig. \ref{fig:Z12}).
(b) View of $x-y$ plane. 6-fold and 4-fold rings are indicated in blue and red, respectively.} \label{fig:per-ext}
\end {figure}

Because a $6/4$ disclination pair constitutes a dislocation, we extend the disclination lines in a parallel fashion
as indicated in Fig. \ref{fig:x-z-view}. 
The 6-fold disclinations run through sequences of $Z14$-coordinated Nb atoms \cite{nelson1983order}, while the
4-fold disclination runs through a sequence of $Z10$-coordinated Fe atoms. At the bottom the three disclinations merge and
annihilate in another $Z13$. This structure has been relaxed using DFT. In Fig. \ref{fig:2a_final} we show the structure can be extended
periodically in both the vertical ($z$) and horizontal ($x$) directions. We leave it as a finite slab in the perpendicular ($y$)
direction as shown in Fig. \ref{fig:2b_final}. Aluminum atoms marked as Al-i have perfect icosahedral coordination. The three
distorted rectangles are drawn surrounding one such atom in Fig. \ref{fig:2a_final}. Also visible in Fig. \ref{fig:2a_final} are additional 6-fold disclination lines running perpendicular to the figure marked with *, and one of the surrounding 6-rings is also shown.
\begin{figure}
\centering
\includegraphics[width=6.5in, height=2.0in]{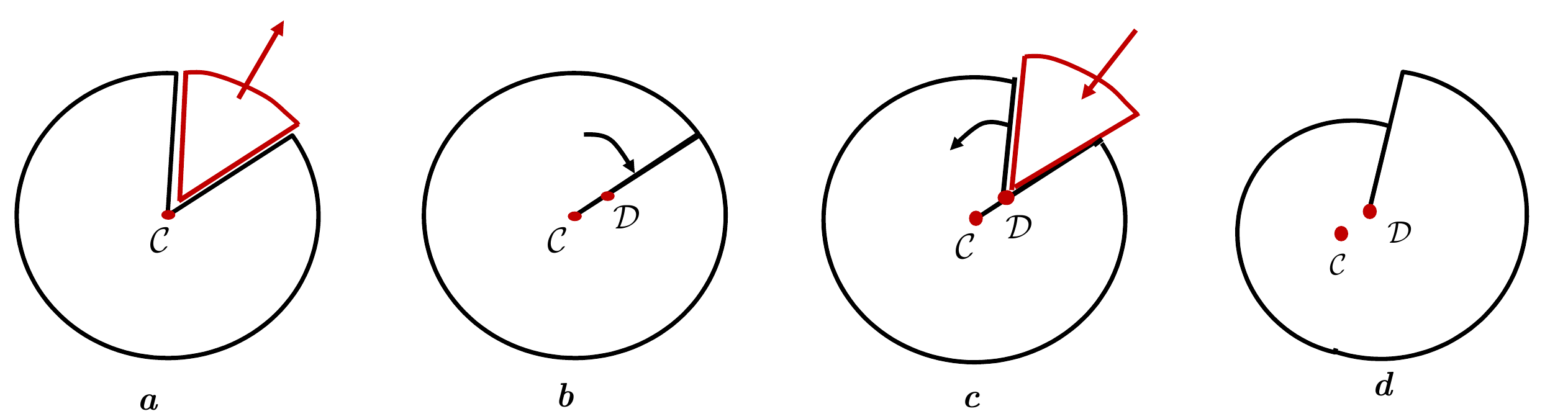}
\caption{A rendering of Eshelby's \cite{eshelby1956continuum} rationalization of a disclination dipole as a dislocation.}
\label{fig:eshelby}
\end{figure}
With regard to the mechanical model of defect dynamics proposed subsequently, we note the very high density of defects that arise naturally in atomic configurations of these metallic glass alloys.

We now adapt Eshelby's \cite{eshelby1956continuum} rationalization for obtaining a dislocation from the combination of two disclinations to the glass dislocation scenario. This is described in Fig. \ref{fig:eshelby}. In Fig. \ref{fig:eshelby}a a wedge is taken out of a stress-free disk, the exposed faces brought together and welded (Fig. \ref{fig:eshelby}b). The body is then allowed to relax (elastically). This gives rise to the stress fields induced by the 5 and 4-fold disclination lines in the glass, when the angle of the wedge is interpreted appropriately. 

\begin{figure}
\centering
\includegraphics[width=6.5in, height=2.5in]{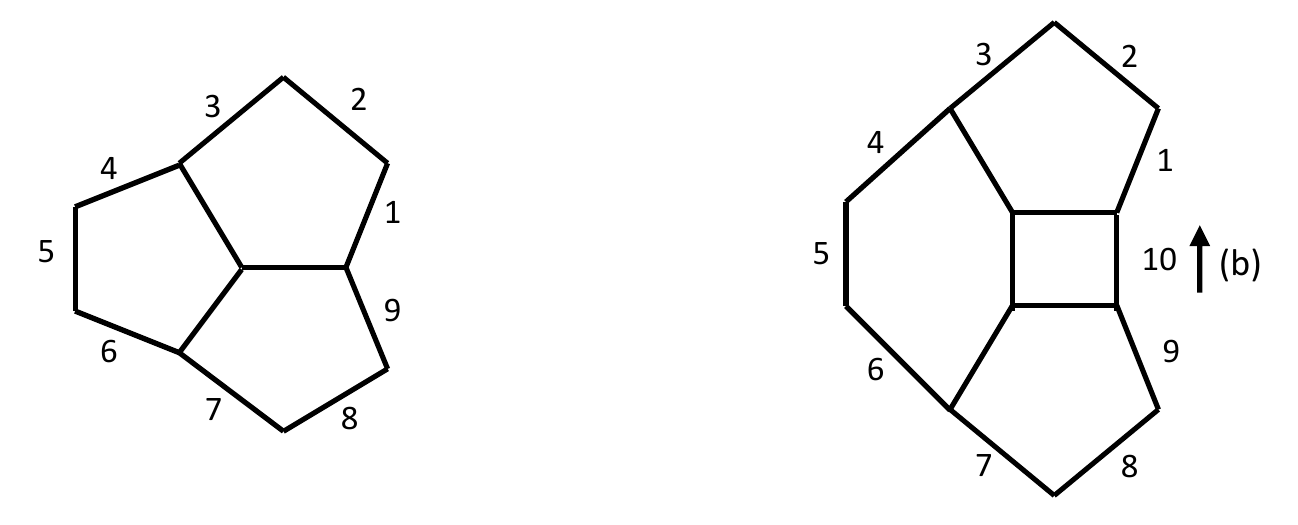}
\caption{Sketch of Burgers vector of a $6/4$ disclination pair, described here in the cofiguration deformed from the pentagonal state.}
\label{fig:burgers}
\end{figure}

Imagine now a stressed pentagonal sea. In that is inserted first a 4-fold disclination as shown in Fig. \ref{fig:eshelby}a,b. Finally, an extra wedge is inserted in this body as in  Fig. \ref{fig:eshelby}c. The apex of the extracted and inserted wedges do not coincide as shown. The inserted wedge is then welded in place and again the body is allowed to relax. This gives the state of stress of a glass edge dislocation. Figure \ref{fig:burgers} shows a sketch motivating how a Burgers vector may be associated with a $6/4$ disclination pair such as those illustrated in the atomistic model shown in Fig. \ref{fig:2b_final}.

Based on the above physical picture, we model the elasticity of the glass as arising from the stress strain response of elementary tetrahedra with unstressed reference edge lengths deduced from the pair-potential in question. This elastic material is thought of as
\begin{itemize}
\item filled with immobile 5-fold disclination lines. Furthermore, 
\item the material contains dislocation lines that are composites of 4-6 disclination dipoles as discussed above. These glass dislocations are considered as mobile. 
\item An excess of 6-fold immobile disclinations may be considered as present as well. 
\end{itemize}
In the next Section we propose a dynamical model for the mechanics of this model of defects in a metallic glass.

\section{A model for the dynamics of glass dislocations and some implications}\label{sec:disloc_theory}

For the sake of simplicity in this first attempt, we consider a model whose stress response is assumed to be linear elastic but one which involves nonlinearity in the evolution of the dislocation density and plastic distortion fields. Such linear elasticity of glasses can be well characterized from DFT calculations \cite{widom1986icosahedral, widom2011elastic}. A model of continuum defect dynamics including the coupling of dislocations and generalized disclinations (g.disclinations) and considerations of unrestricted material and geometric nonlinearity has been developed \cite{acharya2012coupled, acharya2015continuum}. On a refined assessment of the predictions of the current dislocation-only model, such a more general model may very well be necessary in case it is found, on microscopic atomistic grounds, that the assumption of the Burgers vector of a composite $6/4$ defect being constant along its line is not a tenable approximation when measured on the length scales of interest. In fact, since in the metallic glass media under consideration it is the disclination lines that are most readily rationalized based on atomistic configurations/dense particle packings, the general theory may very well be based on g.disclinations alone, with the mobile defects being combinations of specific types of such defects that roughly correspond to dislocations, as discussed earlier. Even so, such a theory is significantly more complicated than the dislocation-only model; therefore, it is reasonable to develop the glass-dislocation model and explore its dynamics with respect to capability in predicting observed behavior of metallic glasses. The following account is a start in that direction.

We first consider the question of designing a model for calculating the \emph{static} stress field of the background of 5-fold disclination lines along with the excess 6-fold disclination lines combined with a prescribed dislocation density. To simplify matters, we consider a (rotational) disclination line as a special case of a defect line that is the terminating curve of a surface of discontinuity of the entire elastic distortion, including strain and rotation. We call such defects as generalized disclinations, or g-disclinations. Then the strength of the g-disclination line may be viewed as the jump in the elastic distortion across the surface of discontinuity. The corresponding g-disclination density field is a third-order tensor field with support on the g-disclination core cylinder, given by the tensor product of the strength per unit core cross-sectional area and the unit tangent vector of the line (for details see \cite{acharya2015continuum}). For multiple disclination lines in the body, the g-disclination density field is simply the sum of the individual density fields, denoted by $\bfPi$. Assume for the moment that the entire spatial distribution of the five-fold and excess six-fold  disclination lines is known and therefore the field $\bfPi$. The g-disclination density field has the property that when acting on a unit normal to a planar surface through a point, it produces the strength per unit area (around the point in question) of all g-disclinations threading the oriented plane being considered; when integrated over any area patch in the body it yields the net strength of the disclination lines threading the area patch.

Similarly, let $\bfalpha$ be the (glass) dislocation density field in the body with the property that when integrated on any area patch it yields the net Burgers vector of the dislocation lines threading the area patch. Then the following set of equations suffices to compute the stress field of this combined collection of dislocation and disclination lines:
\begin{align}\label{eq:stress}\nonumber
curl \, \bfS &= \bfPi  &\epsilon_{ijk}\partial_j S_{mnk} &= \Pi_{mni}\\
div \, \bfS & = {\bf0} &\partial_k S_{mnk} &= 0\\\nonumber
curl \, \bfU^e &= - \bfS:\bfX + \bfalpha & \epsilon_{ijk}\partial_j U^e_{rk} &= -S_{rmn}\epsilon_{mni} + \alpha_{ri}\\\nonumber
div \, \bfT + \bff & = {\bf0} & \partial_j T_{ij} + f_i
 & = 0 \\\nonumber
 \bfT & = \bfC \bfU^e & T_{ij} &= C_{ijkl} U^e_{kl}
\end{align}
with the boundary conditions
\begin{align*}
\bfS\,\bfn = \bf0 \\\nonumber
\bfT \, \bfn = \bft.
\end{align*}
In the above and what follows, all tensor components and spatial partial derivatives are written w.r.t. a rectangular Cartesian coordinate system. $\bfS$ is the elastic 2-distortion field that transmits the effects of the disclination density to the stress $\bfT$, $\bfU^e$ is the elastic distortion tensor, $\bfC$ is the tensor of elastic moduli (with minor and major symmetries), $\bff$ is the body force density, $\bfn$ is the unit normal field on the boundary of the body, and $\bft$ is a statically admissible traction field. It is understood that the dislocation field is augmented beyond the glass dislocation data to ensure that $div \, \bfalpha = div \, \bfS:\bfX$ and that $\bfPi$ is solenoidal. The first two equations of (\ref{eq:stress}) along with the first boundary condition allow for a determination of the elastic 2-distortion arisng due to the disclination density field as source. The third equation represents the fundamental kinematics of the origin of the elastic distortion, sourced by the elastic incompatibilities arising from the presence of the disclination and dislocation densities. The fourth equation is the statement of balance of forces while the fifth is the elastic stress-strain relationship.

It can be shown that with $\bfPi$, $\bfalpha$, $\bff$, and $\bft$ as prescribed data the above set of equations enable the unique determination of $\bfS$ and $\bfT$ and $\bfU^e$ (the latter up to an insignificant spatially uniform skew tensor), under standard positive -definiteness assumptions on the elastic moduli. Linearity and uniqueness ensures that in the absence of the dislocation and disclination density fields and any external loading, the stress vanishes in the body. It can also be shown that the above set of equations recover the classical solutions of individual dislocation and disclination line fields, and when the cores are non-singular, they result in non-singular stress fields with finite total energy in the body.

Given the reality of the glass structure, the background 5-fold disclination density is expected to be so dense that practically it may well be impossible to specify this component of the overall disclination density field in detail. However, it is present and it needs to be accounted for in some way. Our equations above provide the structure for calculating the background stress field of a \emph{stochastic} specification of the 5-fold disclination density field arising from insufficient knowledge of structure.

The considerations above may be considered as setting the initial conditions for stress. We are now interested in equations that model the evolution of the glass dislocation density in the body (i.e. the defect structure evolution) and its coupling to the stress field evolution. We are also interested in the evolution of the total deformation of the field, which is an important observable for our purposes. Because  the motion of individual disclinations require energetically expensive long-range reorientations of matter, we assume that the disclination density field $\bfPi$ does not evolve in time. To this end, we introduce a displacement field, $\bfu$, of the reference configuration on which the initial internal stress field was calculated above. We also introduce a plastic distortion tensor $\bfU^p$ by the relation
\begin{align}\label{eq:Up}
\bfU^p &= grad\, \bfu - \bfU^e & U^p_{ij} &= \del_j u_i - U^e_{ij}.
\end{align}
Physically, the irrotational part of the plastic distortion tensor represents permanent, stress-free deformation taken up by the body (in the absence of displacement boundary conditions preventing such). We now use the tautology that the rate of change of the net Burgers vector content of any area patch is equal to the Burgers vector carried in minus that carried out of the area patch through its boundary due to motion of the dislocations. This motion is represented by a velocity field $\bfV$ associated with the dislocation density field and it can be shown \cite{acharya2011microcanonical} that the above statement translates to the mathematical statement
\begin{align}\label{eq:alpha_evol}
\dot{\bfalpha} &= - curl \, \left( \bfalpha \times \bfV \right) & \dot{\alpha}_{ij} &= - \epsilon_{jmn} \del_m \left( \epsilon_{nrs} \alpha_{ir} V_s \right).
\end{align}
where a superposed dot represents a partial derivative with respect to time. Combining (\ref{eq:alpha_evol}), (\ref{eq:Up}), and (\ref{eq:stress})$_3$ with the observation that $\bfS$ does not evolve in time, we obtain
\begin{equation}\label{eq:Up_evol}
\dot{\bfU}^p = \bfalpha \times \bfV
\end{equation}
through the neglect of a gradient of a vector field, motivated by the physical assumption that all plastic distortion is associated with the motion of dislocations. The equations associated with the displacement field is the balance of linear momentum
\begin{equation}\label{eq:bolm}
div \, \bfT + \bff = \rho \ddot{\bfu},
\end{equation}
where $\rho$ is the mass density and we assume the stress to be a symmetric tensor so that angular momentum balance is automatically satisfied.

A \emph{minimal set of governing equations for the glass dislocation model} then consists of (\ref{eq:bolm}), (\ref{eq:Up_evol}), (\ref{eq:Up}), (\ref{eq:stress})$_5$, with initial conditions on $\bfU^p$ generated from the use of (\ref{eq:stress}) and (\ref{eq:Up}) with $\bfu = \bf0$.

The question now turns to finding some robust guidelines for the choice of the constitutive function $\bfV$ to close the model. For this, we turn to a `mechanical' version of the second law of thermodynamics and require that the rate of working of the external forces on the body always exceed or equal the sum of the rate of change of kinetic energy and the rate at which energy is stored in the body, i.e.
\begin{align}\label{dissipation}\nonumber
{\cal{D}} &:= \int_B \bff \cdot \dot{\bfu} \,dv + \int_{\del B} (\bfT \bfn) \cdot \dot{\bfu} \, da - \frac{d}{dt} \int_B \frac{1}{2} \rho |\dot{\bfu}|^2 \, dv   - \int_B \dot{\psi} \, dv \\
 & = \int_B \bfT : grad \, \dot{\bfu} \, dv - \int_B \dot{\psi} \, dv \geq 0,
\end{align}
where $\psi$ is the free-energy density per unit volume on the body $B$. Assuming a free-energy dependence of the form (cf. \cite{zhang2015single})
\begin{equation*}
\psi = \psi \left( \bfU^e, \bfU^p, \bfalpha \right)\footnote{Obviously, our considerations do not preclude the free-energy density not having a dependence on some of its arguments displayed above, if eventually deemed appropriate on physical grounds. One way of maintaining compact dislocation cores while working with the practically useful assumption of a convex dependence of the free energy density on the total displacement gradient requires a non-convex dependence of on a field representing localized atomic motions from a well-defined (possibly stressed) atomic configuration); our $\bfU^p$ is a representative of such a physical notion.},
\end{equation*}
and using the field equations (and ignoring boundary contributions to the dissipation for simplicity), we have
\begin{align*}
\mathcal{D} & = \int_B \left( \bfT - \del_{\bfU^e} \psi \right): grad \, \dot{\bfu} \, dv + \int_B \mathcal{F} \cdot \bfV \, dv & \\
\mathcal{F} & := \bfX \left( \del_{\bfU^e} \psi - \del_{\bfU^p} \psi + curl \, \bfalpha \right)^T \bfalpha & \mathcal{F}_i = \epsilon_{ijk} \left( \del_{U^e_{rj}} \psi - \del_{U^p_{rj}} \psi + \epsilon_{jmn} \del_m \alpha_{rn} \right) \alpha_{rk}.
\end{align*}
We think of \emph{$\mathcal{F}$ as the driving force for the glass dislocation velocity} and invoke the reversible elastic response
\[
\bfT =  \del_{\bfU^e} \psi.
\]
Next, we define
\begin{align*}
\bfa & = \frac{1}{3}tr(\bfT) \bfX \bfalpha &  a_i & = \frac{1}{3} T_{mm} \epsilon_{ijk} \alpha_{jk} & \bfb & = \mathcal{F} - \bfa \\
\bfd & = \bfb - \left( \bfb \cdot \frac{\bfa}{|\bfa|} \right) \frac{\bfa}{|\bfa|} & \bfp & = \bfa - \left( \bfa \cdot \frac{\bfb}{|\bfb|} \right) \frac{\bfb}{|\bfb|}\\
\bfd & = \bfb \ \ \mbox{if} \ \ \bfa = {\bf0} & \bfp & = \bfa \ \  \mbox{if} \ \  \bfb = {\bf0}
\end{align*}
and a general class of constitutive equations for $\bfV$ of the form
\begin{equation}\label{velocity}
\bfV = s_d \left(\bfa, \bfb \right)\, \frac{\bfd}{|\bfd|} + s_p (\bfa, \bfb) \,\frac{\bfp}{|\bfp|},
\end{equation}
where $s_p$ and $s_d$ are non-negative real-valued functions of their arguments. Then
\[
\mathcal{D} = \int_B \mathcal{F} \cdot \bfV \, dv = \int_B (\bfa + \bfb) \cdot \left( s_d (\bfa, \bfb) \, \frac{\bfd}{|\bfd|} + s_p (\bfa, \bfb)\, \frac{\bfp}{|\bfp|} \right) \, dv \geq 0
\]
(by Pythagoras' theorem).

We note that the structure of the model and the minimal requirement of consistency with a (reduced form) of the second law of thermodynamics implies that the dislocation velocity and the consequent \emph{plastic flow is dependent on the pressure}. $s_p = 0$ and independence of $s_d$ on its first argument yields the pressure independent part of the dislocation velocity.

Next we consider the dilatational component of the plastic strain rate,
\begin{align*}
tr \, \dot{\bfU^p}  & = tr (\bfalpha \times \bfV) = \hat{\bfa} \cdot \bfV & \dot{U}^p_{ss} & = \hat{a}_s V_s & \hat{\bfa} &= \bfX \bfalpha & \hat{a}_s &= \epsilon_{sir} \alpha_{ir}.
\end{align*}
Defining $\mathsf{p} = - \frac{1}{3} tr(\bfT)$ and noting that $\bfd$ is orthogonal to $\hat{\bfa}$ we obtain
\[
tr\,\dot{\bfU}^p = -\mathsf{p} \, s_p  \left [ \hat{\bfa} \cdot \hat{\bfa} - \frac{ \left( \hat{\bfa} \cdot \bfb \right)^2 } {\bfb \cdot \bfb} \right]
\]
(with the second term in the parenthesis being $0$ when $\bfb = \bf0$). Thus, the \emph{plastic dilatation is in general non-vanishing} and its sign is governed by the sign of the pressure $\mathsf{p}$. It can be observed that if the dislocation state at a point is of edge character then plastic dilatation arises due to climb motions as physically expected, with no dilatation associated with screw components.

It is known that the evolution equation for the dislocation density (\ref{eq:alpha_evol}) is fundamentally of transport type \cite{acharya2001model} and is capable of predicting longitudinal extension of shear bands as shown in \cite{zhang2015single}. This is an important problem for models of shear banding in metallic glasses \cite{greer2013shear}. In \cite{zhang2015single} it is also shown that such models of dislocation mechanics show a threshold stress for the onset of dislocation motion, despite being translationally-invariant. We believe that such a feature coupled with the strong spatial heterogeneity of the background stress field in a glass will provide a fundamental basis for the prediction of macroscopic yielding in our model. Finally, plastic deformation in glasses, even when monitored in atomistic simulations, is not observed to arise from the more-or-less uniform sweeping motion of dislocation lines but as sudden shearings of clusters of atoms. We conjecture that such behavior can be reconciled with the dislocation picture by noting that because of the heterogeneous background stress field, it is possible that at any given time only segments of a dislocation loop undergo motion, the rest being pinned (especially by the transverse defect lines that thread them, as shown in Fig. \ref{fig:2a_final}).

\section*{Acknowledgments}
A.A acknowledges support in part from grants NSF-CMMI-1435624, NSF-DMS-1434734, and ARO W911NF-15-1-0239. M.W acknowledges support from DOE grant DE-SC0014506.
\bibliographystyle{alpha}\bibliography{glass_bib}

\end{document}